\documentclass[showpacs,prl,aps,superscriptaddress,twocolumn]{revtex4}
\usepackage{mathrsfs}
\usepackage{array}
\usepackage{amsmath}
\usepackage{graphicx}
\usepackage{amstext}
\usepackage{amsfonts}
\usepackage{bm}
\usepackage{epstopdf}

\usepackage{color}
\usepackage{hyperref}
\usepackage{calc}
\begin{document}
\title{Exact homogeneous master equation \\for open quantum systems incorporating initial correlations}
\author{Pei-Yun Yang}
\affiliation{Department of Physics and Center for Quantum
Information Science, National Cheng Kung University, Tainan 70101,
Taiwan}
\author{Wei-Min Zhang}
\email{wzhang@mail.ncku.edu.tw}
\affiliation{Department of Physics
and Center for Quantum Information Science, National Cheng Kung
University, Tainan 70101, Taiwan}
\begin{abstract}
We show that the exact master equation incorporating
initial correlations for open quantum systems, within the
Nakajima-Zwanzig operator-projection method, is a homegenous master
equation for the reduced density matrix. We also using the quantum Langevin equation to derive
explicitly the exact master equation incorporating initial correlations for a large class of bosonic 
and fermionic open quantum systems described by the Fano-Anderson Hamiltonian, and the
resulting master equation is homogenous. We find that the
effects of the initial correlations can be fully embedded into the
fluctuation dynamics through the exact homogenous master equation.
Also a generalized nonequilibrium fluctuation-dissipation
theorem incorporating various initial correlations is obtained.
\end{abstract}

 \maketitle
Realistic systems in nature have inevitable interactions with its
surrounding environments. When such interactions are not negligible,
the systems must be treated as an open system \cite{Feynman63}. 
Many physical, chemical and biological systems, and all kinds of quantum devices 
developed in recent years for nano and quantum technologies, should be
considered as open quantum systems. In contrast to an isolated
quantum system, whose dynamics is governed by the Schr\"{o}dinger
equation, the dynamical evolution of an open quantum system can be
described by the master equation \cite{Breuer08,Weiss08}. Formally,
the exact master equation for arbitrary open system is obtainable
through the operator-projection method which was initially proposed
by Nakajima \cite{Nakajima58} and Zwanzig \cite{Zwanzig60}. When there are initial
correlations between the system and its environment, it was claimed 
\cite{Breuer08} that, the Nakajima-Zwanzig (NZ) equation
becomes inhomogeneous. However, if a maser equation for open systems is 
inhomogeneous, it may violate the trace-preserving property of the density matrix and 
thereby induce unphysical states to the open system dynamics. In this work, we prove 
that the exact master equation is always homogeneous for any open quantum system
incorporating arbitrary initial correlations. 


In fact, nonequilibrium dynamics of open systems incorporating
initial correlations is a long-standing problem in mesoscopic physics and
also in the foundation of statistical mechanics \cite{Grabert88,Haug2008}. In the past two decades,
investigation of quantum nonequilibrium dynamics has mainly been focused
on steady-state phenomena \cite{Imry2002,Blanter2000}, where
initial correlations are usually considered to be not essential due to the memory
loss and therefore are often ignored. Recent experimental developments
allow one to measure transient quantum dynamics of open systems,
in particular, for nano-scale quantum devices and various atomic and molecular
structure-based biological systems \cite{Lu03422,Bylander05361,
Gustavsson08152101,Lee2007,Engle2007}. In the transient nonequilibrium 
regime, initial correlations could provide important information on the formation 
and evolution of a large number of different physical objects. 
In this work, we will also derive explicitly the exact homogeneous
master equation for a large class of open systems incorporating initial
correlations. 

\textit{1. NZ master equation is homogenous.}
The NZ equation was originally proposed to
describe the dynamics of a subspace of interest,  projected out from
the whole Hilbert space ${\cal E}$ of a large system \cite{Nakajima58,Zwanzig60}.
It begins with a decomposition of the Hilbert space into a relevant part
plus an irrelevant part,
\begin{align}
\rho_{tot}(t) = {\cal P}\rho_{tot}(t) + {\cal Q}\rho_{tot}(t) ,
\end{align}
where ${\cal P}$  and ${\cal Q}$  are the two projection operators that satisfy the
projection operator properties:  ${\cal P}+ {\cal Q}=I$, ${\cal P}^2={\cal P}$,
${\cal Q}^2={\cal Q}$, and ${\cal P}{\cal Q}={\cal Q}{\cal P} =0$. From
the quantum Liouvillian  equation of the system,  $\frac{d}{dt}\rho_{tot}(t)
=- i[H_{tot}, \rho_{tot}(t)]\equiv {\cal L}(t)\rho_{tot}(t)$, one can formally
derive the following NZ equation for the relevant subspace of the system,
\begin{align}
\frac{d}{dt}{\cal P}\rho_{tot}(t) &= {\cal P}{\cal L}(t){\cal P}\rho_{tot}(t) +
{\cal P}{\cal L}(t){\cal G}(t,t_0){\cal Q}\rho_{tot}(t_0) \notag \\
&+\int^t_{t_0}d\tau {\cal P}{\cal L}(t){\cal G}(t,\tau){\cal Q}{\cal L}(\tau){\cal P}
\rho_{tot}(\tau)  ,  \label{NZme}
\end{align}
where  ${\cal G}(t,\tau)$ is an evolution operator given by
\begin{align}
  {\cal G}(t,\tau) = {\cal T} \exp \Big\{ \int^t_\tau ds {\cal Q}{\cal L}(s)\Big\},
\end{align}
and ${\cal T}$ is the time-ordering operator. 
In Ref.~\cite{Breuer08}, it is claimed that the second term in Eq.~(\ref{NZme}), 
which is obtained from the initial system-environment correlations, is an inhomogeneous 
term to the master equation of open systems. Only when there is no initial 
system-environment correlation, the second term vanishes, and the corresponding master 
equation becomes a homogeneous master equation \cite{Breuer08}. 
We will first show that this claim is incorrect.

To be explicit, let us focus on the large system 
consisting of a subsystem of interest (refereed to the
principal system \cite{Feynman63}, i.e. the open system hereafter)
coupling to the rest as its environment, then by specifying the projection
operator,
\begin{align}
{\cal P}\rho_{tot}\equiv {\rm Tr}_E[\rho_{tot}]\otimes {\rm Tr}_S[\rho_{tot}]
=\rho_S \otimes\rho_E,
\end{align}
one projects the entire Hilbert space of the
system into the subspace written apparently as direct products
of the density matrices of the open system with its environment.
Notice that ${\rm Tr}_E$ and ${\rm Tr}_S$ are traces over separately
the environmental states and the system states. Thus,
$\rho_S={\rm Tr}_E[\rho_{tot}]$ and $\rho_E={\rm Tr}_S[\rho_{tot}]$
are the corresponding reduced density matrices describing
respectively quantum states of the open system and the environment that
mutually carry the entanglement information between the system and the environment.
The left part of the Hilbert space is ${\cal Q}\rho_{tot}=\rho_{tot}-\rho_S \otimes\rho_E$.
By taking a trace over all the environmental states on the both sides of Eq.~(\ref{NZme}),
the NZ equation is reduced to the master equation describing
the time evolution of the reduced density matrix $\rho_S(t)$ for the open system.

Now, we can show that the second term in Eq.~(\ref{NZme}) is not an inhomogeneous
term to the master equation for the reduced density matrix $\rho_S(t)$.
Note that the second term in Eq.~(\ref{NZme}), ${\cal P}{\cal L}(t)
{\cal G}(t,t_0){\cal Q}\rho_{tot}(t_0) $ must be a density matrix belonging to the
same subspace of ${\cal P}\rho_{tot}(t)$
due to the project operator ${\cal P}$. Then the trace over the environment states
on the second term in Eq.~(\ref{NZme}) gives a
reduced density matrix $\rho'_S(t)$ which could be (and usually is)
different from  $\rho_S(t)$:
\begin{align}
{\rm Tr}_E[{\cal P}{\cal L}(t){\cal G}(t,t_0){\cal Q}\rho_{tot}(t_0)]
\equiv \rho'_S(t) \neq \rho_S(t) .  \label{icd}
\end{align}
However, $\rho'_S(t)$ and $\rho_S(t)$ are two different reduced density matrices of the open system,
it is {\it always} possible  to connect $\rho'_S(t)$ with $\rho_S(t)$ by superoperators, i.e.
\begin{align}
\rho'_S(t) \longrightarrow \sum_{ij} \chi_{ij}(t) A_j \rho_S(t) B_i,  \label{icds}
\end{align}
where $\chi_{ij}(t)$ is in general a time-dependent coefficient that depends
on the initial system-environment correlations, and
$A_i, B_j$ are operators of the open system. More specifically, 
these superoperators can be written with the 
following symmetric and trace-preserving form:
\begin{align}
\rho'_S(t) \longrightarrow \sum_{ij} \chi_{ij}(t) \big[& A_j \rho_S(t) B_i + B_i \rho_S(t) A_j \notag \\
& - A_jB_i\rho_S(t) -  \rho_S(t)B_iA_j \big],  \tag{\ref{icds}$^\prime$}
\end{align}
Consequently,  it shows that the second term
in Eq.~(\ref{NZme})  can always be reduced to an homogeneous term to
the master equation.

In fact, the second and the third terms in the NZ equation (\ref{NZme}) have  
the same operator structure. This can be seen explicitly by writing ${\cal L}(\tau){\cal P}
\rho_{tot}(\tau) \equiv \rho''_{tot}(\tau) $, as another state of the total system.
Then the integration function of the non-local time integral in the third term
of Eq.~(\ref{NZme}) can be rewritten as
\begin{align}
{\cal P}{\cal L}(t){\cal G}(t,\tau){\cal Q}{\cal L}(\tau){\cal P}
\rho_{tot}(\tau) \!=\! {\cal P}{\cal L}(t){\cal G}(t,\tau){\cal Q}\rho''_{tot}(\tau) , \label{NZ3t}
\end{align}
which shows the same operator structure as the second term,
${\cal P}{\cal L}(t){\cal G}(t,t_0){\cal Q}\rho_{tot}(t_0)$ in Eq.~(\ref{NZme}).
Thus, after taken the trace over all the environment states, the second term in
the NZ equation (\ref{NZme}) must produce a homogenous term to the 
master equation 
as does the third term. 
In other words, initial correlations {\it cannot} produce an inhomogeneous
term to the master equation for open systems. 
The claim in Ref.~\cite{Breuer08} that initial correlations result in an 
inhomogenous master equation cannot be correct.

\textit{2. Derivation of the exact homogenous master equation for the Fano-Anderson model}.
On the other hand, using cumulant expansions \cite{Kubo63} rather than the NZ
operator-projection methods, it was formally shown
that the reduced density matrix $\rho_S(t)$ obeys an exact 
homogeneous evolution equation when the initial system-environment
correlations are considered \cite{Royer96}.  Recently, explicit exact
homogeneous master equations are derived for the pure
dephasing model of a two-level system \cite{Ban2009}, for the nano-cavity
system in photonic crystals \cite{Tan2011}, and for the nano-structured electronic
systems in mesoscopic physics \cite{Yang2015}, where initial correlations
between the system and its reservoirs are taken into account.  Here we shall derive
further an exact homogeneous master equation incorporating
initial system-environment correlations for a large class of open systems
investigated widely in condensed mater physics and atomic physics.

These open systems we concerned here are described by the Fano-Anderson
Hamiltonian, 
\begin{align}
H^{FA} (t) =& H_S(t) + H_E(t) + H_{SE}(t) \notag \\
=& \sum_{ij} \varepsilon_{ij} (t)a^\dag_i a_j \notag +
\!\sum_{\alpha k} \epsilon_{\alpha k}(t)
 b^\dag_{\alpha k} b_{\alpha k}  \notag \\
 & +\!\sum_{\alpha ik}\!\!\Big(V_{i \alpha k}(t)a^\dag_i b_{\alpha k}
 \!+\! V^*_{i \alpha k}(t) b^\dag_{\alpha k} a_i\!\Big)  ,  \label{FAH}
\end{align}
as it was originally introduced by Anderson \cite{Anderson1961} and
Fano \cite{Fano1961} independently for investigating impurity electrons
in solid-state physics and discrete states embedded in a continuum in
atomic physics, respectively.  In Eq.~(\ref{FAH}), we let all the
energy levels and couplings be time-dependent because these
parameters can be manipulated experimentally through the new development of
nanotechnologies in the investigation of nonequilibrium dynamics.  In the past
two decades, the Fano-Anderson Hamiltonian has often been
used as a starting point to investigate various nonequilibrium transport phenomena
and decoherence dynamics in semiconductor nanoelectronics and
spintronics \cite{Mahan2000,Haug2008, Imry2002}.   The  Fano-Anderson Hamiltonian
was also referred as the Lee-Friedrichs Hamiltonian \cite{Friedrichs1948,
Lee1954} in atomic and molecular physics, nuclear physics
and  scalar field theory with the physcial relevance of Fano resonances,
Dicke effect and superradiance, etc.

Let us consider the first case for Eq.~(\ref{FAH}), in which the total system is initially in
equilibrium,  which is called the partition-free scheme in the literature \cite{Cini805887,Stef04195318}:
\begin{align}
\rho_{tot}(t_0)=\frac{1}{Z}e^{-\beta(H^{FA}(t_0)-\mu N)}.
\end{align}
Obviously, this initial state contains initial system-environment correlations.
In other words, ${\cal Q}\rho_{tot}(t_0) \neq 0$ in the NZ project method so that  
the second term in Eq.~(\ref{NZme}), regarded as an inhomogeneous term in 
\cite{Breuer08}, does not vanish.
Meantime, by experimentally tuning on the time-dependent parameters in Eq.~(\ref{FAH})
through external fields, the total system becomes nonequilibrium, and the time
evolution of the total density matrix is determined by
\begin{align}
\rho_{tot}(t) = U(t,t_0)\rho_{tot}(t_0)U^\dag(t,t_0),  \label{tdm}
\end{align}
where $U(t,t_0) = {\cal
T}\!\exp\big\{\!\!-\!\!i\!\int^t_{t_0}\!\!d\tau H^{FA}(\tau)\big\}$
is the evolution operator of the total system. Because $H^{FA}(t)$
in Eq.~(\ref{FAH}) has a bilinear form of the particle creation and
annihilation operators, the total density matrix $\rho_{tot}(t)$
will remain as a Gaussian function all the time. Taking the trace
over the environment states (e.g.~by path-integral method with the
Gaussian integral), the reduced density matrix $\rho_S(t)={\rm
Tr}_E[\rho_{tot}(t)]$ is still a Gaussian function of the system
variables only, with a prefactor which is only a function of time.
Then the time-derivative of the reduced density matrix must be
proportional to the same Gaussian function. Consequently, the
resulting master equation must be homogeneous.

Actually, using the Feynman-Vernon influence functional approach \cite{Feynman63},
we have previously derived the exact homogenous master equation of the reduced density matrix
$\rho_S(t)$ from Eq.~(\ref{tdm}),
for both boson and fermion systems in the absence of
initial system-environment correlation \cite{Tu2008, Jin2010,Lei2012,Zhang2012},
where the initial state of the system and the environment 
is assumed to be decoupled \cite{Leggett1987}.
In this exact master equation formalism, the environment-induced energy-level
renormalization, the particle dissipations, and thermal fluctuations, due to the
influence of the environment, are naturally incorporated into the time-dependent
dissipation and fluctuations coefficients in the exact master equation. These
damping and fluctuation coefficient are determined exactly and nonperturbatively
by the Schwinger-Keldysh nonequilibrium Green's functions in many-body
systems \cite{Schwinger1961,Kadanoff1962}. By solving these
nonequilibrium Green's functions, we investigated nonperturbatively various environment-induced
non-Markovian dynamics through the time evolution of open quantum systems, in the
absence of initial system-environment correlations \cite{Zhang2012}.

Now, to derive the exact master equation incorporating the initial system-environment
correlations, we utilize the exact quantum Langevin equation from Eq.~(\ref{FAH})
\cite{Tan2011,Yang14115411,Yang2015},
\begin{align}
\frac{d}{dt}a_i(t)\! =\!-\!i\sum_j \bm{\varepsilon}_{ij}(t) a_j(t)
\!-\!\!\sum_{j}\!\!\int_{t_0}^t \!\! \! d\tau
\bm{g}_{ij}(t,\tau)a_j(\tau) \!+ \!f_i(t) ,  \label{qle}
\end{align}
where
\begin{subequations}
\begin{align}
&\bm{g}_{ij}(t,\tau) = \sum_{\alpha k}V_{i\alpha k}(t)V^*_{j\alpha
k}(\tau)e^{-i\!\int_{\tau}^{t}\!\!\epsilon_{\alpha k}(\tau_1)d\tau_1} , \label{nlik}\\
&f_i(t)= -i\sum_{\alpha k} V_{i\alpha k}(t)
b_{\alpha k}(t_0)e^{-i\!\!\int_{t_0}^t \!\!\bm{\epsilon}_{\alpha k}
(\tau_1) d\tau_1}, \label{lat}
\end{align}
\end{subequations}
which is obtained by eliminating {\it exactly} all the environmental degrees of
freedom through the Heisenberg equation of motion, and it is valid
for arbitrary initial state of the system and the environment.
Because of the linearity, the general solution of Eq.~(\ref{qle}) has the form
\cite{Tan2011,Yang14115411,Yang2015}
\begin{align}
a_i(t) = \sum_{j}\bm u_{ij}(t,t_0)a_j(t_0) + F_i(t) ,    \label{als}
\end{align}
where $\bm u_{ij}(t,t_0)=\langle [a_i(t), a^\dag_j(t_0)] \rangle$ 
is the propagating Green function of the particles in the system,
which carries all the information of the dissipation dynamics of the system induced by
the tunneling coupligs between the system and the environment, and $F_i(t)$ is the
fluctuation function resulting from the noise force $f_i(t)$ of the environment, 
see Eqs.~(\ref{qle}) and (\ref{lat}).

It is easy to show from Eqs.~(\ref{qle}) and (\ref{als}) that the propagating Green function
obeys the Dyson equation,
\begin{align}
&\frac{d}{dt}\bm{u}(t,t_0) \!+\!
i\bm{\varepsilon}(t)\bm{u}(t,t_0)\!+\!\! \int_{t_0}^{t}\!\!\!
\!d\tau\bm{g} (t,\tau)\bm{u}(\tau\!,t_0) \!= \!0,   \label{ue}
\end{align}
subjected to the initial condition $\bm{u}(t_0,t_0) = \bm 1$, where
the time non-local integral kernel, $\bm{g} (t,\tau)$ given by Eq.~(\ref{nlik}),
describes the probability amplitude of a particle tunneling from the
system into the environment at time $\tau$, propagating in the
environment within the time $\tau$ to $t$, and then tunneling back
into the system at time $t$. Thus, the $\tau$-integration from the
very initial time $t_0$ to the present time $t$ through this non-local time
integral kernel in Eq.~(\ref{ue}) records all the historical evolution of the
particle dissipated from the system into the environment. This characterizes
the microscopic picture of the non-Markovian memory processes
\cite{Zhang2012}.
The fluctuation function $F_i(t)$, a contribution from the inhomogeneous color-noise
force $f_i(t)$ in  the quantum Langevin equation (\ref{qle}), is given explicitly by
\begin{align}
&F_i(t) \!= \!-i \!\sum_{\alpha k j} \!\!\int_{t_0}^t \!\!\!d\tau\big[\bm u_{ij}(t,\tau)V_{j\alpha k}(\tau)
e^{-i\!\!\int_{t_0}^\tau \!\!\bm{\epsilon}_{\alpha k} (\tau_1) d\tau_1} \!\big]c_{\alpha k}(t_0),\label{nf}
\end{align}
which describes particles tunneling from the environment
into the system at time $\tau$, and then propagating
to the level $i$ during the time from $\tau$ to $t$.
This provides the underlying fluctuation dynamics induced by particles initially in
the environment. The initial operator-dependence in Eqs.~(\ref{als})
and (\ref{nf}), proportional to $a_i(t_0)$ and $c_{\alpha k}(t_0)$
respectively, fully incorporate all the initial correlations of the
total system.

Combining the dissipation and fluctuation dynamics together through
the above exact quantum Langevin equation, we obtain the exact homogenous
master equation for the reduced density matrix of
the system incorporating various initial correlations:
\begin{align}
& {d\rho_S(t)\over dt} =  -i\big[H'_S(t),\rho_S(t)\big] \notag  + \sum_{ij}\Big\{ 
\bm{\gamma}_{ij}(t) {\cal D}(a_j,a^\dag_i) \\
&~+\!\widetilde{\bm{\gamma}}_{ij}(t){\cal F}(a_j,a^\dag_i) +\! \overline{\gamma}_{ij}(t) 
{\cal F}(a^\dag_j,a^\dag_i)+\! \overline{\gamma}^\dag_{ij}(t){\cal F}(a_j,a_i) \Big\} ,
\label{ME}
\end{align}
The renormalized Hamiltonian $H'_S=\sum_{ij}
\bm{\varepsilon}'_{ij}(t)a^\dag_ia_j$, and the dissipation and fluctuation superoperators:
\begin{subequations}
\begin{align}
{\cal D}(a_j,a^\dag_i)&\!=\!2a_j\rho_S(t) a_i^{\dag}\!-\!a_i^{\dag}a_j\rho_S(t)  \!-\! \rho_S(t) a_i^{\dag}a_j  , \\
{\cal F}(a_j,a^\dag_i)&\!=\!a^\dag_i\rho_S(t)a_j \!\pm\!a_j\rho_S(t)a_i^\dag
 \!\mp\! a_i^{\dag}a_j\rho_S(t) \!-\! \rho_S(t) a_ja_i^{\dag} .
\end{align}
\end{subequations}
where the up and low signs of $\pm$ correspond to both the system and reservoirs
consisting of bosons or fermions. 
The renormalized energy $\bm{\varepsilon}'_{ij}(t)$, the dissipation and fluctuation coefficients,  
$\bm{\gamma}_{ij}(t)$, $\widetilde{\bm{\gamma}}_{ij}(t)$ and
$\overline{\bm{\gamma}}_{ij}(t)$ are fully determined by the following relations,
\begin{subequations}
\label{ecoff1}
\begin{align}
\bm{\varepsilon}'_{ij}(t)= &
\frac{i}{2}\big[\dot{\bm{u}}(t,t_0)\bm{u}^{-1}(t,t_0) - {\rm
H.c.}\big]_{ij} ,
\\
\bm{\gamma}_{ij}(t)= & -\frac{1}{2}\big[\dot{\bm{u}}(t,t_0)\bm{u}^{-1}(t,t_0) + {\rm H.c.}\big]_{ij},
\\
\widetilde{\bm{\gamma}}_{ij}(t)\!=& \dot{\bm{v}}_{ij}(t,t)-[\dot{\bm{u}}(t,t_0)\bm{u}^{-1}(t,t_0)\bm{v}(t,t)\!+\! {\rm H.c.}]_{ij} 
, \\
\overline{\bm{\gamma}}_{ij}(t)\!=& \mp\frac{1}{2}\dot{\bm{\nu}}_{ij}(t,t)\pm
\frac{1}{2}[\dot{\bm{u}}(t,t_0)\bm{u}^{-1}(t,t_0)\bm{\nu}(t,t)\!+\!
{\rm T}]_{ij}  \label{gamt}
\end{align}
\end{subequations}
where $\bm{u}(t,t_0)$ is the propagating Green function of
Eq.~(\ref{ue}), and $\bm{v}(t,t)$ and $\bm{\nu}(t,t)$ are the
particle-particle and particle-pair correlation Green functions in
nonequilibrium many-body systems,
\begin{subequations}
\label{cgfn}
\begin{align}
&\bm{v}(\tau,t) \!=\!\! \int_{t_0}^{\tau} \!\!\! \!d\tau_1
\!\! \int_{t_0}^t \!\!\! d\tau_2
\bm{u}(\tau,\tau_1)\widetilde{\bm{g}}(\tau_1,\tau_2)
\bm{u}^{\dag}(t,\tau_2), \label{ve}  \\
&\bm{\nu}(\tau,t) \!=\!\! \int_{t_0}^{\tau} \!\!\! \!d\tau_1 \!\!
\int_{t_0}^t \!\!\! d\tau_2
\bm{u}(\tau,\tau_1)\overline{\bm{g}}(\tau_1,\tau_2)
\bm{u}^T(t,\tau_2), \label{nue}
\end{align}
\end{subequations}
which are directly proportional to the initial correlations through the non-local time integral kernels,
and $\widetilde{\bm{g}}(\tau_1, \tau_2)$ and $\overline{\bm{g}}(\tau_1, \tau_2)$,
\begin{subequations}
\label{tnl}
\begin{align}
\widetilde{\bm{g}}_{ij}(\tau_1,\tau_2) &=\widetilde{\bm{g}}^{sb}_{
ij}(\tau_1,\tau_2)+\widetilde{\bm{g}}^{bb}_{ij}(\tau_1,\tau_2),   \label{wtg} \\
\overline{\bm{g}}_{ij}(\tau_1, \tau_2) &=\overline{\bm{g}}^{sb}_{ij}(\tau_1,\tau_2)+
\overline{\bm{g}}^{bb}_{ij}(\tau_1,\tau_2),   \label{otg}
\end{align}
\end{subequations}
with
\begin{subequations}
\label{corr}
\begin{align}
\widetilde{\bm{g}}^{sb}_{ij}(\tau_1, \tau_2)=&\!-\!2i\!\sum_{\alpha k}\!\!\Big[
V_{i\alpha k}(\tau_1) e^{-i\!\int_{t_0}^{\tau_1}\! \!\epsilon_{\alpha k}(\tau)d\tau}\!
\delta(\tau_2\!-\!t_0)  \notag \\
\times & \! \langle a_{j}^\dag(t_0)b_{\alpha k}(t_0)\rangle
\!-\! {\rm H.c}(i\!\leftrightarrow\! j, \tau_1\!\leftrightarrow \!\tau_2)\Big], \\
\widetilde{\bm{g}}^{bb}_{ij}(\tau_1,\tau_2)=&\!\! \sum_{\alpha k \alpha' k'}\!\!\!
V_{i\alpha k}(\tau_1)e^{-i\!\int_{t_0}^{\tau_1}\!\!\epsilon_{\alpha
k}(\tau)d\tau} V^*_{j\alpha' k'}(\tau_2)  \notag \\
& \times \!\! e^{i\!\int_{t_0}^{\tau_2}\!\!\epsilon_{\alpha'
k'}(\tau)d\tau}\!\langle b_{\alpha' k'}^\dag(t_0)b_{\alpha
k}(t_0)\rangle ,  \\
\overline{\bm{g}}^{sb}_{ij}(\tau_1,\tau_2)=&\!-\!2i\!\sum_{\alpha
k}\!\!\Big[ V_{i\alpha k}(\tau_1) e^{-i\!\int_{t_0}^{\tau_1}\!
\!\epsilon_{\alpha k}(\tau)
d\tau}\!\delta(\tau_2\!-\!t_0) \notag \\
&\times \!\! \langle a_{j}(t_0)b_{\alpha k}(t_0)\rangle \! \pm \!
(i\!\leftrightarrow\! j,
\tau_1\!\leftrightarrow \!\tau_2)\Big], \\
\overline{\bm{g}}^{bb}_{ij}(\tau_1,\tau_2)=&-\!\!\!\!\sum_{\alpha k
\alpha' k'}\!\!\! V_{i\alpha
k}(\tau_1)e^{-i\!\int_{t_0}^{\tau_1}\!\!\epsilon_{\alpha
k}(\tau)d\tau} V_{j\alpha' k'}(\tau_2) \notag \\
& \times \!\! e^{-i\!\int_{t_0}^{\tau_2}\!\!\epsilon_{\alpha'
k'}(\tau)d\tau}\!\langle b_{\alpha' k'}(t_0)b_{\alpha
k}(t_0)\rangle.
\end{align}
\end{subequations}
The integral kernels $\widetilde{\bm{g}}^{sb}(\tau_1,\tau_2)$ and
$\overline{\bm{g}}^{sb}(\tau_1,\tau_2)$ are proportional to initial particle-particle
and particle-pair correlations between the system and the environment,
and $\widetilde{\bm{g}}^{bb}(\tau_1,\tau_2)$ and $\overline{\bm{g}}^{bb}
(\tau_1,\tau_2)$ are associated respectively with initial particle-particle correlations and
particle-pair correlations in the environment. The particle-pair correlations
can exist when the system and the environment are initially in, for examples, 
a superconducting state (for electron systems) \cite{Lai} or a squeezed state 
(for photon systems) \cite{Tan2011}.
In fact, because the Hamiltonian (\ref{FAH}) is bilinear, other higher-order
correlations can be uniquely determined in terms of these basic nonequilibrium
two-time Green functions, $\bm u(t, \tau)$, $\bm v(\tau, t)$ and $\bm \nu (\tau, t)$.

\textit{3. Unification with the influence functional approach and the
nonequilibrium Green function technique.}
When there are no initial particle-pair correlations, namely
$\langle a_{i}(t_0)b_{\alpha k}(t_0)\rangle=0$ and $\langle
b_{\alpha' k'}(t_0) b_{\alpha k}(t_0)\rangle=0$, we will have
$\overline{\bm{g}}(\tau_1,\tau_2)=0$. As a result, $\bm{\nu}(t,t)=0$
and also $\overline{\bm \gamma}(t)=0$ so that the exact master
equation (\ref{ME}) is reduced to the one incorporating initial
particle-particle correlations only that we derived very recently
\cite{Yang2015}. If the initial system-environment correlations are
also negligible,  $\langle a_{j}^\dag(t_0) b_{\alpha k}(t_0)\rangle
\simeq 0$ and $\langle b^\dag_{\alpha k}(t_0)a_{i}(t_0)\rangle
\simeq 0$, namely, the system is initially decoupled from the
environment \cite{Leggett1987}, and meanwhile the environment is initially in the
thermal equilibrium, i.e. $\langle b_{\alpha' k'}^\dag(t_0)b_{\alpha
k}(t_0)\rangle = f(\varepsilon_{\alpha k}, T)\delta_{\alpha
\alpha'}\delta_{kk'}$,  then
\begin{align}
\widetilde{\bm{g}}_{ij}(\tau ,\tau')\!=\!&\sum_{\alpha k}
V_{i\alpha k}(\tau)V^*_{j\alpha k}(\tau') e^{-i\! \int_{\tau'}^{\tau}\! \epsilon_{\alpha
k}(\tau_1)d\tau_1} \!f(\varepsilon_{\alpha k}, T)    \label{cni}
\end{align}
where $f(\varepsilon_{\alpha k}, T)= \frac{1}{e^{\beta (\varepsilon_{\alpha k}-\mu_\alpha )} \mp 1}$
is the initial Bose-Einstein (Fermi-Dirac) distribution function at temperature $T$ for
bosonic (fermionic) environments. The resulting exact master equation
goes to the master equation of Eq.~(2) in Ref.~[\onlinecite{Zhang2012}] for the case of the
decoupled initial system-environment states that can be derived directly \cite{Tu2008,Jin2010,Lei2012,Zhang2012} 
using the Feynman-Vernon influence functional approach \cite{Feynman63}.
However, in all the cases, the exact master equation for open systems are always homogenous,
no matter if there are initial correlations or not.

As a self-consistent check, we show further that both the inhomogenous quantum Langevin 
equation (\ref{qle}) and the homogenous master equation (\ref{ME}) give the same nonequilibrium 
lesser Green functions:
\begin{subequations}
\label{cgf}
\begin{align}
\bm{G}^{<}_{ij}(\tau,t)&= i\langle a^\dag_j(t) a_i(\tau)\rangle
\notag\\&
=\big[\bm{u}(\tau,t_0)\bm{G}^<(t_0,t_0)\bm{u}^{\dag}(t,t_0)+i\bm{v}(\tau,t)\big]_{ij} , \\
 \overline{\bm{G}}^{<}_{ij}(\tau,t)&= i\langle a_j(t) a_i(\tau)\rangle
\notag\\&
=\big[\bm{u}(\tau,t_0)\overline{\bm{G}}^{<}(t_0,t_0)\bm{u}^T(t,t_0)+i\bm{\nu}(\tau,t)\big]_{ij}
,
\end{align}
\end{subequations}
This shows that the inhomogeneity in physical observables induced by
initial correlations can be fully taken into account in the exact
homogeneous master equation.  In fact, the Keldysh's correlation
Green function $\bm v(t, t)$ and the generalized particle-pair
correlation $\bm \nu(t,t)$ in Eq.~(\ref{cgf}), given explicitly by
Eq.~(\ref{cgfn}), provide indeed the generalized nonequilibrium
quantum fluctuation-dissipation theorem in the time domain. 
In the weak system-environment coupling regime,
the nonequilibrium quantum fluctuation-dissipation theorem will be
reduced to the equilibrium fluctuation-dissipation theorem in the
steady-state limit \cite{Kubo1966} (a simple proof can be
found in the Supplemental Material of Ref.~\onlinecite{Zhang2012}).
Furthermore, in the high temperature limit, the equilibrium
fluctuation-dissipation theorem is simply reduced to the Einstein's
fluctuation-dissipation theorem \cite{Einstein1905} which can also
be derived from the classical Langevin equation \cite{Langevin1908}.
This provides a consistent check of the theory. 

In conclusion, we show that the Nakajima-Zwanzig master equation in terms
of the reduced density matrix for open quantum systems incorporating initial 
correlations is always a homogenous equation. We also derived explicitly, 
through the exact quantum Langevin equation, the exact homogenous master equation 
incorporating initial correlations. This master equation can describe the noneqilibrium dynamics for 
a large class of open quantum systems based on the Fano-Anderson Hamiltonian 
which has widely been used in atomic physics,
condensed matter physics and various nano-scale systems in the past decades.
The resulting exact homogenous master equation recovers the master equation derived from 
Feynman-Vernon influence-functional in the absence of the initial correlations as a special limit. 
The dissipation and fluctuation dynamics, embedded in the time-dependent coefficients
of the exact homogenous master equation, which take into account all initial correlation 
effects, are fully determined by the generalized Schwinger-Keldysh nonequlibrium
Green functions.  Thus, for the very first time, by connecting the  Nakajima-Zwanzig method, 
the quantum Langevin equation method, the Feynman-Vernon influence functional approach 
and the Schwinger-Keldysh nonequlibrium Green function technique together, we develop 
the most general exact homogenous master equation for a large class of open quantum systems.
We expect that this underlying master equation incorporating initial correlations would find
more applications for the investigations of nonequilibrium dynamics not only in physics,
but also in chemical and biological systems through the exploration of general dissipation
and fluctuation dynamics.

\section*{Acknowledgment}
This research is supported by the Ministry of Science and Technology of ROC under
Contract No. NSC-102-2112-M-006-016-MY3, and the National Center for Theoretical
Science of Taiwan. It is also supported in part by the Headquarters
of University Advancement at the National Cheng Kung University,
which is sponsored by the Ministry of Education of ROC.

\end{document}